# Sphaleron-like processes in a realistic heat bath

A. Krasnitz[a][*] and R. Potting[b]

[a]IPS, ETH-Zentrum, CH-8092 Zürich, Switzerland

[b]Universidade do Algarve, Unidade de Ciências Exactas e Humanas,
Campus de Gambelas, 8000 Faro, Portugal

We measure the diffusion rate of Chern-Simons number in the (1+1)-dimensional Abelian Higgs model interacting with a realistic heat bath for temperatures between 1/13 and 2/3 times the sphaleron energy. It is found that the measured rate is close to that predicted by the sphaleron approximation at the lower end of the temperature range considered but falls at least an order of magnitude short of the sphaleron estimate at the upper end of that range. We show numerically that the sphaleron approximation breaks down as soon as the gauge-invariant two-point function yields correlation length close to the sphaleron size.

In the study of thermally induced transitions over energy barriers in field theories numerical real-time simulations are often the only computational tool reliable in a wide range of temperatures. Determination of the corresponding transition rate in a simulation should include its proper averaging over the canonical ensemble in the phase space of a system in question. If such averaging is done by immersing the system in a heat bath, the latter should be coupled in a way that leaves the bulk dynamical properties of the system intact. This is achieved by allowing interaction with the heat bath only at boundaries. The choice of the heat bath is further restricted by requiring that its dynamical properties will resemble as closely as possible those of the system itself. We are then led to the following realistic heat bath (RHB) recipe: linearize the field theory to be studied in the vicinity of a vacuum and use the resulting system as a boundary-coupled heat bath [2]. In our earlier work [1] we verified that this construction works as intended in the simplest case of one-component field in one dimension. Here we report on applying the RHB technique to the study of sphaleron transitions in the Abelian Higgs model (AHM) in one spatial dimension [3]. Our interest in these processes is due to their analogy to baryon-number violating transitions in the Standard Model, believed to have played an important role in setting the baryon number of the Universe to its present value [4]. In particular, AHM possesses an infinite number of degenerate vacua separated by finite energy barriers. The sphaleron solution corresponding to the barrier saddle point is known analytically [5–7,9].

The lattice AHM Hamiltonian in the temporal gauge is in suitably chosen units [5]

$$H = \frac{1}{2a}\sum_j [\xi E_{j,j+1}^2 + |\pi_j|^2 + \frac{a^2}{2}(|\phi_j|^2 - 1)^2 + |\phi_{j+1} - e^{iaA_{j,j+1}}\phi_j|^2], \quad (1)$$

where $\pi_j$ and $E_{j,j+1}$ are canonically conjugate momenta of the complex scalar (Higgs) field $\phi_j$, and the spatial component of the gauge field $A_{j,j+1}$, respectively. The lattice spacing is $a$. The Hamiltonian equations of motion following from (1) are easy to integrate numerically if a Cartesian representation is used for the Higgs field. For this reason we use them for real-time evolution in the bulk of an open gauge-Higgs system. We have verified that the Gauss' law constraint

$$\frac{1}{a}(E_{j,j+1} - E_{j-1,j}) = \operatorname{Im}(\pi_j \phi_j^*) \quad (2)$$

is maintained to a high precision for very long integration times.

On the other hand, RHB is most easily constructed in terms of gauge-invariant variables. In order to obtain a suitable set of variables the Higgs field is rewritten in polar coordinates: $\phi_j = \rho_j \exp(i\alpha_j)$. We then define $b_{j,j+1} = \alpha_{j+1} - \alpha_j - aA_{j,j+1}$, $\epsilon_{j,j+1} = \frac{1}{a}E_{j,j+1}$ and introduce canonical momentum $\pi_j^\rho$ for $\rho_j$. The corresponding canonical momentum for $\alpha_j$ is expressed in terms of $\epsilon$ variables using (2). It is easy to see that $\rho_j, \pi_j^\rho$ and $\epsilon_{j,j+1}, b_{j,j+1}$ form two pairs of canonical variables whose dynamics is governed by

$$H' = \frac{a}{2}\sum_j\left(\xi\epsilon_{j,j+1}^2 + \left(\frac{\epsilon_{j,j+1}-\epsilon_{j-1,j}}{a\rho_j}\right)^2\right)$$
$$+\frac{a}{2}\sum_j\left(\left(\frac{\pi_j^\rho}{a}\right)^2 + +\frac{1}{2}\sum_j(\rho_j^2-1)^2\right)$$
$$+\frac{1}{a}\sum_j\left((\rho_j^2 - \rho_j\rho_{j+1}\cos b_{j,j+1})\right), \quad (3)$$

obtained from polar-coordinate form of (1) by substituting (2).

---

[*]Speaker at the conference.

As a heat bath we take AHM linearized in the vicinity of one of its gauge-equivalent vacua, which, as is well known, is a system of two free fields: the radial Higgs field $\varrho$ and the gauge field $\varepsilon$ whose masses are $\sqrt{2}$ and $\sqrt{\xi}$, respectively. Those are coupled at the boundary site of the AHM ((1) or (3)) each to its interacting counterpart, i.e. $\varrho$ to $\rho$, and $\varepsilon$ to $\epsilon$. The field equations at the boundary are modified compared to those in the bulk by introducing terms to account (a) for the heat-bath reaction to the boundary motion, and (b) for thermal fluctuations of the heat bath. The reader is referred to Ref. [1] for details of the heat-bath construction and numerical implementation. In order to be able to transform from polar coordinates, used for RHB, to Cartesian ones, used in the bulk, we need to keep track of the angular variable $\alpha$ of the boundary Higgs field. This is done with the help of Gauss' law which for the linearized system gives

$$\dot{\alpha}_j = \frac{1}{a}(E_{j,j+1} - E_{j-1,j}). \quad (4)$$

The focal point of our study was determining the diffusion rate of the Chern-Simons number $N_{\rm CS} \equiv (2\pi)^{-1}\int A(x)dx$. Following Ref. [5], we extracted the rate from $\Delta_{\rm CS}(t)$, the average squared deviation of $N_{\rm CS}$ for a lag $t$. For sufficiently long $t$ a random-walk behavior sets in:

$$\Delta_{\rm CS}(t) = \Gamma L t, \quad (5)$$

where $\Gamma$ is the transition rate per unit length.

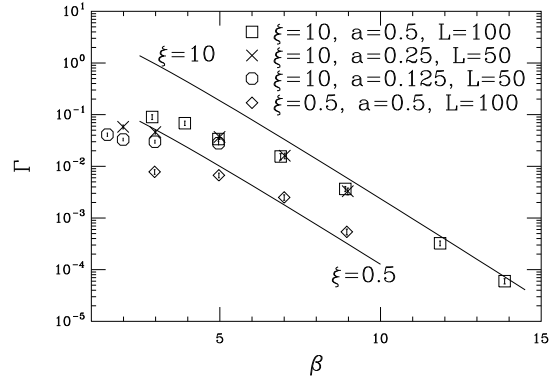

Figure 1. Temperature dependence of the transition rate $\Gamma$. The solid curves correspond to (6), with the value of $\xi$ indicated near each curve.

Our rate measurements are to be compared to the prediction of Ref. [6] derived under the assumption that $\Gamma$ can be identified with the rate of vacuum-to-vacuum transitions, and that the latter involve configurations closely resembling the sphaleron:

$$\Gamma \propto \sqrt{\beta}\exp(-\beta E_{\rm sph}), \quad (6)$$

where $E_{\rm sph}$ is the sphaleron energy. As Figure 1 clearly shows, the values of $\Gamma$ at low temperature slowly approach those given by the sphaleron approximation. The two nearly coincide at the lowest temperature considered, $\beta = 14$. Note also that the lattice spacing dependence of the rate is absent in the low-temperature regime, in agreement with the findings of earlier work [7,5]. But the most notable feature of our results is their dramatic departure downwards from the sphaleron prediction starting at about $\beta = 5$. Our measurements indicate that the rate practically does not grow in a range of temperatures to the left of $\beta = 5$ point. This range extends to higher and higher temperatures as we reduce the lattice spacing.

It is natural to ask why this strong lagging of the measured rate behind the one given by (6) begins in the vicinity of $\beta = 5$. To this end recall the underlying assumption of (6): Chern-Simons number diffusion is dominated by evolution of configurations resembling the vacuum into those resembling the zero-temperature sphaleron. It is clear that the $N_{\rm CS}$ diffusion can only be described in these terms as long as the Higgs field is correlated on a length scale larger than the sphaleron size ($2\sqrt{2}$). The corresponding correlation length $\lambda$ can be found by measuring a gauge-invariant two-point function [8]

$$C_{jl} = \phi_j^* \phi_l \exp\left(-ia \sum_{k=j}^{l-1} A_{k,k+1}\right). \quad (7)$$

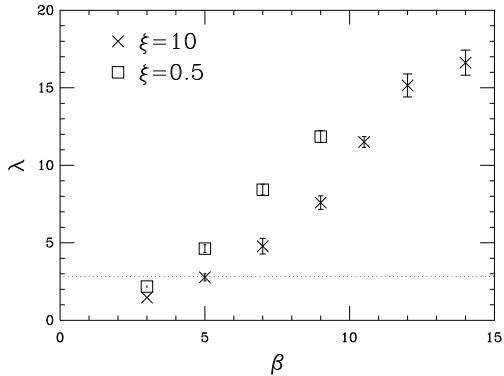

Figure 2. Temperature dependence of the correlation length deduced from (7). The sphaleron size is shown by the dashed line for comparison.

Figure 2 shows the values of $\lambda$ obtained by fitting $\langle C_{jl}\rangle$ to const $\times \exp(-|j-l|/\lambda)$. As expected, the breakdown of the sphaleron approximation for the rate occurs as $\lambda$ becomes smaller than the sphaleron size. Our measurements of $\lambda$ are also consistent with the $a$ dependence of the rate at high temperatures. For example, at $\beta = 3$ $\lambda = 1.47$ for $\xi = 10$, only about 3 times larger than the lattice spacing $a = 0.5$. Hence the field strongly fluctuates at length scales comparable to the lattice spacing.

At this point it is unclear what causes the sharp slowdown of the rate growth at the high-temperature end of our measurement range. We cannot exclude a possibility that at temperatures in question crossing the barrier between two neighboring vacua in the configuration space of the model [9] in close vicinity of the sphaleron saddle point is still strongly preferred energetically, but is already suppressed entropically. It is usually assumed [9–12] that at temperatures above the sphaleron energy the rate grows like a power of the temperature. It could be that what we observe at $\beta \leq 5$ is a crossover from the exponential to power-law behavior of the rate. It is not clear, however, that at such a crossover the rate should stop growing as our measurements seem to indicate. We hope to resolve this puzzling situation by rate measurements at still higher temperatures, as well as smaller lattice spacings.